\documentclass[twocolumn,prc]{revtex4}
\usepackage{epsfig}
\usepackage{amssymb}

\begin{document}
\bibliographystyle{revtex}

\title{ Elastic p$-^{\text{3}}$He and n$-^{\text{3}}$H scattering with
  two- and three-body forces }
\author{ B.\ Pfitzinger }
\author{ H.~M.~Hofmann }
\affiliation{Institut f{\"u}r Theoretische Physik III,
   University of Erlangen-N{\"u}rnberg, Staudtstra\ss{}e 7,
  D 91058 Erlangen, Germany}
\author{ G.~M.~Hale }
\affiliation{Theoretical Division, Los Alamos National Laboratory,
  Los Alamos, NM 87545, USA}

\begin{abstract}
  We report on a microscopic calculation of $n-^3$H and $p-^3$He
  scattering employing the Argonne $v_{18}$ and $v_8'$ nucleon-nucleon
  potentials with and without additional three-nucleon force. An
  $R$-matrix analysis of the $p-^3$He and $n-^3$H scattering data is
  presented. Comparisons are made for the phase shifts and a selection
  of measurements in both scattering systems. Differences between our
  calculation and the $R$-matrix results or the experimental
  data can be attributed to only two partial waves ($^3P_0$ and
  $^3P_2$). We find the effect of the Urbana IX and the Texas-Los
  Alamos three-nucleon forces on the phase shifts to be negligible.
\end{abstract}
\maketitle

\section*{Introduction}

It is well known that realistic nucleon-nucleon ($NN$) forces cannot
reproduce the $\rm ^3H$ and $\rm ^3$He binding energies. Three-nucleon
interactions (TNIs) are added to give the necessary small corrections
but they still fail to reproduce certain properties of the three
nucleon system, most notably the $A_{\rm y}$ analyzing power in $Nd$
scattering \cite{AY}.  Yet the 30\% deviation of $A_{\rm y}$ can be
resolved by tiny changes in the $Nd$ scattering phase shifts (on the
order of $0.1$ degrees \cite{PD_3MEV, KIEVSKY_ND, nd_psa}).
Furthermore very many operators can contribute to a TNI and the lack
of stringent conditions in the three-nucleon system on the structure
of the TNI makes its application to other systems desirable.  In
\cite{HE4} it was shown that although a realistic $NN$ force can generally
reproduce the $^4$He system, there remain differences,
most notably in the analyzing powers.  The intensely studied $\rm
^4He$ system \cite{TILLEY_A4} is unfortunately very difficult to
describe due to the many resonances and the $\rm^4He$ bound state.
Therefore we investigate the much simpler systems $p-^3$He and
$n-^3$H where data exist in the energy range of interest.

We organize the paper in the following way: The next section contains
a description of the $R$-matrix analysis of the $p-^3$He and $n-^3$H data.
After that we discuss the Resonating Group Model (RGM) calculation and
the model space used. Then we compare $R$-matrix and RGM results of
the phase shifts for various model spaces, and demonstrate the
differences for a selection of typical observables. Finally we will
discuss the effect of TNIs on the scattering.

\section{$R$-Matrix Analysis}

The $R$-matrix analysis of the $T=1$ part of the $A=4$ system began
many years ago with an analysis of $p+^3$He scattering data below 20
MeV incident proton energy \cite{HALE_UNPUB}.  Only one solution
described all the data included, namely the one with $\delta(^3P_1) >
\delta(^1P_1)$, giving a $\chi^2$ per degree-of-freedom value of
about 1.23.  Later, this solution was reflected to the $n+^3$H system
\cite{NT_RMAT}, using a simple energy shift to correct for the
short-ranged Coulomb differences between $^4$Li and $^4$H.  This shift
of the $R$-matrix eigenenergies ($\Delta E_\lambda=-0.86$ MeV) for the
$^4$Li system was adjusted by hand to reproduce approximately the $n-t$
total cross section measurement of Phillips, et al.\ \cite{NT_TOTAL}.

More recently, we added to the $p+^3$He analysis analyzing-power and
spin-correlation data measured at energies between 4 and 10 MeV by Alley
and Knutson \cite{HE3_KNUTSON}.  These high-precision data made small
changes in the phase shifts, but did not alter the qualitative nature of
the solution, shown as the solid curves in Figs. 1-5. With these data we
achieve a $\chi^2$ per degree-of-freedom of 1.27.

Finally, we combined the $p+^3$He data with the $n-t$ total cross
sections of ref. \cite{NT_TOTAL} and fitted both reactions
simultaneously.  The single energy shift used earlier did not give a
particularly good fit to the total cross sections, so we allowed the
low-lying eigenenergies in the $^4$H system to adjust separately from
those in the $^4$Li system, while keeping the reduced-width
amplitudes the same in both systems.  This resulted in a $\chi^2$ per
degree-of-freedom of 1.59.  Most of the increase came from the fit to
the $n-t$ total cross sections, which have uncertainties on the order
of 0.2\%.  The $\chi^2$ per point of the fit to the $p+^3$He data
increased only from 1.22 to 1.24.  The analysis is based on 1447 data
points having proton energies between $1.01$ and $19.7$ MeV and neutron
energies between 0.06 and 20.06 MeV, and allows a maximal orbital angular
momentum $L_{\rm max} = 4$.  The $^4$Li resonance energies still agree
with those of \cite{TILLEY_A4} which also contains a brief description of
the $R$-matrix method.

\section{RGM and model space}

We use the Resonating Group Model \cite{RRGM, RRGM_VIEWEG, RRGM_TANG}
to compute the scattering in the $\rm ^4$H and $\rm ^4$Li systems
using the Kohn-Hulth\'en variational principle \cite{KOHN}. The main
technical problem is the evaluation of the many-body matrix elements
in coordinate space. The restriction to a Gaussian basis for the
radial dependencies of the wave function allows for a fast and
efficient calculation of the individual matrix elements \cite{RRGM,
  RRGM_TANG}. However, to use these techniques the potentials must
also be given in terms of Gaussians. In this work we use suitably
parametrized versions of the Argonne $v_{18}$ \cite{AV18} and $v_{8}'$
\cite{AV8P_U8_U9} $NN$ potentials and the Urbana IX \cite{AV8P_U8_U9}
and Texas - Los Alamos \cite{TLA} TNIs.

The inclusion of an additional TNI requires an order of magnitude more
computing power than the realistic $NN$ forces alone. It is therefore
very fortunate that enough data exists at low energies for the
comparatively simple isospin $T=1$ systems $\rm ^4$H and $\rm ^4$Li to
allow for a comparison between calculation and the experimental data
or the $R$-matrix analysis thereof.

In the $^4$Li system we use a model space with three two-fragment
channels, namely the $p - ^3$He, the $^2$H$ - (pp)$ and the
$^2$H$(S=0) - (pp)$ channels. The last two are an approximation to
the three- and four-body breakup channels that cannot in practice
be treated within the RGM. The $\rm ^4$Li is treated as four clusters
in the framework of the RGM to allow for the required internal orbital
angular momenta of $\rm ^3$He or $\rm ^2$H.

For the scattering calculation we include the $S$, $P$ and some of the
$D$ wave contributions to the $J^\pi = 0^+, 1^+, 0^-, 1^- \text{ and }
2^-$ channels. From the $R$-matrix analysis these channels are known
to give essentially the experimental data.  The full wave function for these
channels contains over 100 different spin and orbital angular momentum
configurations, hence it is too complicated to be given in detail. To
give an impression of the model space we will describe the important
structures of $\rm ^3$He as used in the present work.

The dominant spin $S=1/2$ configurations of $\rm ^3$He include those
without angular momentum and with two $D$ waves coupled to total
angular momentum $L=1$. $S=3/2$ in turn occurs together with a single
$D$ wave on each of the Jacobi coordinates or two $D$ waves coupled to
$L=2$. Each of these configurations uses a set of one to three
Gaussians whose width parameters were obtained by a non-linear
optimization using a genetic algorithm \cite{CWGEN}. In this small,
29-dimensional model space we still achieve -6.37 MeV binding energy,
an {\em rms} radius of 1.78 fm and a $D$ state probability of 7.7\%
for the $\rm ^3$He using A$v_{18}$. This must be compared to
$-6.92$MeV known from Faddeev calculations \cite{NOGGA_FAD} whereas we
find $-6.88$MeV in a rather large model space.

This representation of $\rm ^3$He, together with a $S=1/2$ and $L=2$
configuration (excited state) and the $(pn) - (pp)$ fragmentations, form
the basis of our $p - ^3$He scattering calculation.  Once the
fragment wave functions are fixed the scattering problem is solved
with our RGM code relying on the Kohn-Hulth\'en variational principle
\cite{KOHN}:
$$\delta \big( \langle \Psi_t | H-E | \Psi_t \rangle -
\frac{1}{2}a_{ll}\big) =0,
$$
where $a_{ij}$ denotes the reactance matrix.

The model space described above (consisting of two to four physical
scattering channels for each $J^\pi$) is by no means sufficient to
find reasonable results. So-called distortion or pseudo-inelastic
channels \cite{RRGM_TANG} have to be added to improve the description
of the wave function within the interaction region. Accordingly, the
distortion channels have no asymptotic part.

For practical purposes it is obvious to reuse some of the already
calculated matrix elements as additional distortion channels. In that
way we include all the positive parity states of the three-nucleon
subsystem with $J^\pi_3 \le 5/2^+$ in our calculation. However, it was
recently pointed out by A.\ Fonseca \cite{NT_FONSECA} that states
having a negative parity $J_3^-$ in the three-nucleon fragment
increase the $n -^3$H cross section noteably. Therefore we also
added the appropriate distortion channels in a similar complexity as
in the $J_3^+$ case to our calculation, thereby doubling the size of
the model space.

\section{Partial wave analysis}

\begin{figure*}
  \noindent
  \epsfig{file=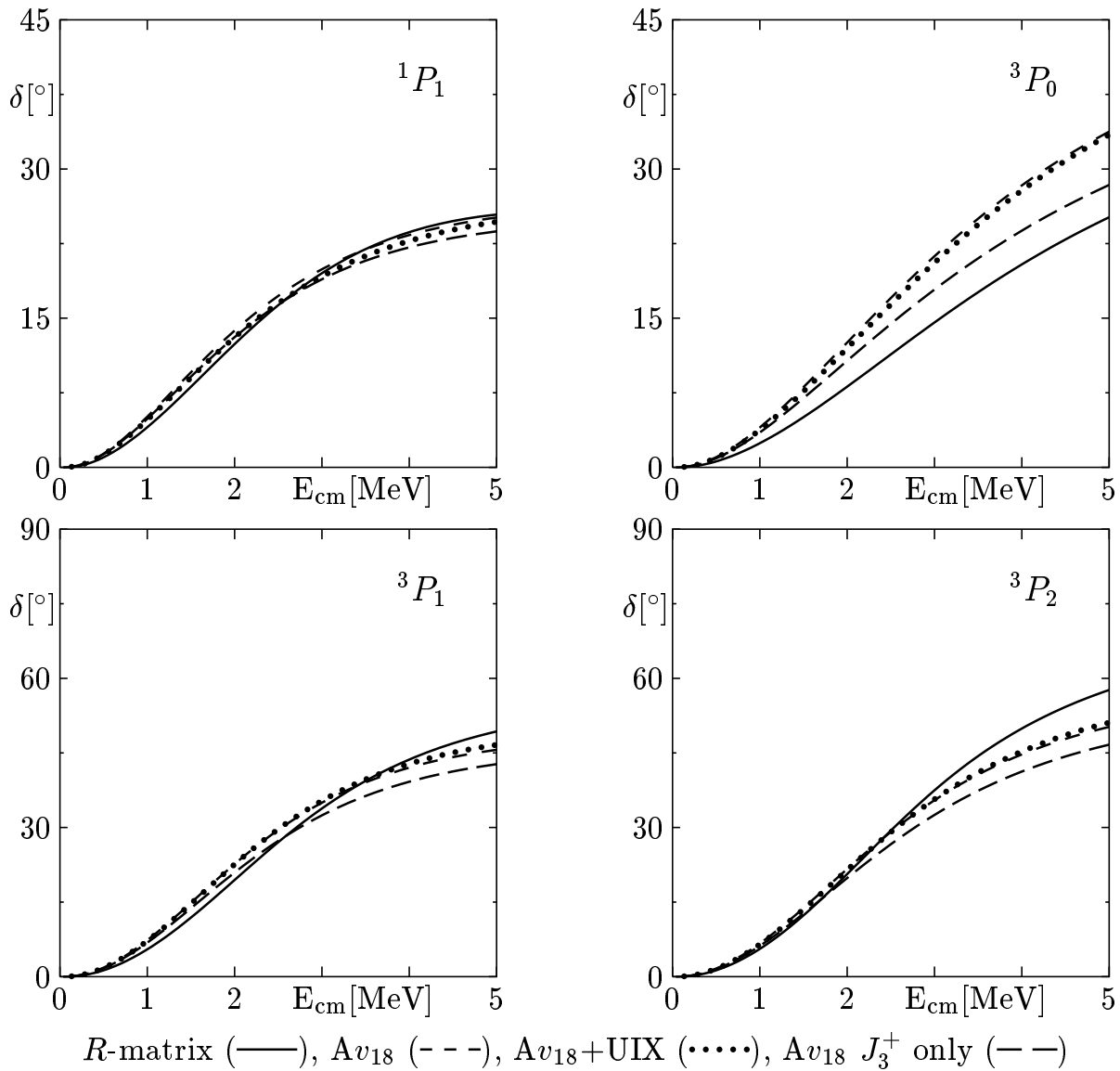, width=0.9\textwidth}
  \caption{ \label{phase_figure} $p - ^3$He $P$ wave scattering phase
    shifts.   RGM calculations (A$v_{18}$, dashed line, restricted to
    postive parity in the three-nucleon subsystem, long dashed, and
    A$v_{18}$    + Urbana IX, dotted) compared to the $R$-matrix analysis
    (solid line).}
\end{figure*}

The $R$-matrix analysis of the $T=1$ part of the $4N$ system 
described above represents the currently available $p-^3$He and
$n-t$ data very well, as the solid lines in figs.\ \ref{he3_diff},
\ref{ay} and \ref{nt_total_xsec} show. In addition it provides
elastic scattering phase shifts (solid line in figs.\
\ref{phase_figure} and \ref{S_NT}) that in turn suffice to describe
the experimental data. Therefore we take these phase shifts as
benchmarks against which to test our calculations.

In a calculation of elastic $p - ^3$He scattering using the full
model space described above (including both the $J_3^+$ and $J_3^-$
distortion channels) we find for both the Argonne $v_{18}$ and $v_8'$
(not shown) interactions in general phase shifts very similar to those
given by the $R$-matrix analysis (see figs.\ \ref{phase_figure} and
\ref{S_NT}). The $S$-waves are negative due to the underlying
Pauli-forbidden states, whereas all the $P$-waves are positve. The
$J_3^-$ part increases the model space and therefore reduces the
repulsion in the $S$-waves and increases the (attractive) interaction
in the $P$-waves.  Hence, all phase shifts have to become more
positive. The $^1S_0$ and $^3S_1$ phase shifts (see fig.~\ref{S_NT})
depend only weakly on the $J_3^-$ part of the model space (long dashed
line as compared to dashed line), because the central terms of the
$NN$-interaction can only connect to a few states of the increased model
space.  The $^3P_0$ and $^3P_2$ phase shifts deviate markedly from the
$R$-matrix results (see fig.~\ref{phase_figure}).  Without the $J_3^-$
components (long dashed line) the major difference between our
calculation and the $R$-matrix results was the $^3P_2$ matrix element
being too small, whereas the $^3P_0$ results were very close.  The
additional $J_3^-$ distortion channels reduce by half the difference for
$^3P_2$, but also raise the $^3P_0$ phase considerably beyond its
$R$-matrix values.

It has already been pointed out previously \cite{HE4} that the
calculated $^3P_2$ matrix element in the $\rm ^4$He system is too
small (then using the realistic Bonn $NN$ interaction \cite{BONNFIT}).
It was shown that the manual change of $^3P_2$ to its $R$-matrix value
removed most of the discrepancies in the description of the
analyzing power. The $J_3^-$ part of the $p-^3$He model space in
the present calculation acts mainly on the $^3P_0$ and $^3P_2$
channels increasing both of them by several degrees (fig.\ 
\ref{phase_figure}). However, the splitting of $^3P_2 - {^3P}_0$
remains considerably too small, see fig.\ \ref{phase_splitting} where
the deviations of the calaculated $p-^3$He phase shifts from the
$R$-matrix values and the $^3P_2-^3P_0$ splitting are shown.  We note
in passing that distortion channels as described above were not
included in \cite{HE4}.

\section{Comparison with data}

\begin{figure}
  \noindent
  \begin{center}
    \epsfig{file=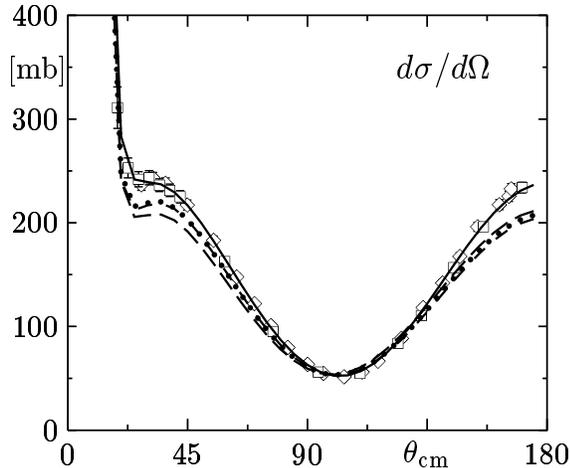, width=0.9\columnwidth}
  \end{center}
  \caption{ \label{he3_diff} $p - ^3$He differential cross
    section at $E_{\rm cm} = 4.1 \text{ MeV}$. The experimental data is from
    \cite{HE3_CLEGG} ($\lozenge$) and \cite{HE3_PHYSREV} ($\square$), the
    long-dashed line denotes the results for the A$v_8'$+TLA interaction,
    other lines as in fig.\ \ref{phase_figure}.
    }
\end{figure}

The increase of the $^3P_2$ matrix element due to the $J_3^-$
components suffices to describe the total and differential cross
sections (figures \ref{he3_diff} and \ref{nt_total_xsec}) in both the
$n -^3$H and $p - ^3$He scattering. Again, the improvement is
due to the larger $^3P_2$ matrix element, the effect of $^3P_0$ being
too large on the cross sections is negligible (on the order of 1\%)
because of the small statistical weight.
\begin{figure}
  \noindent
  \begin{center}
    \epsfig{file=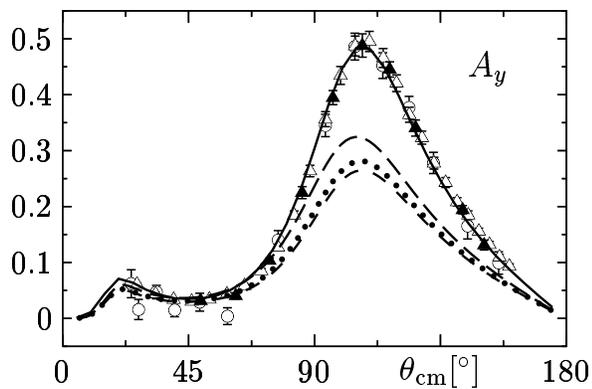, width=0.9\columnwidth}
  \end{center}
  \caption{ \label{ay} $p - ^3$He analyzing power $A_{\rm y}$ at
    $E_{\rm cm} = 4.1 \text{ MeV}$. The experimental data is from
    \cite{HE3_NPA126} ($\circ$) and \cite{HE3_KNUTSON}
    ($\vartriangle$, $\blacktriangle$), the lines are as in fig.\ 
    \ref{phase_figure}.  }
\end{figure}

There remain deviations of the differential cross section at small and
large scattering angles as well as in the low energy behavior of the
total cross section. In the first case, we attribute the differences at
large scattering angles to be an effect of the model space lacking
higher partial waves, and our calculation agrees with the results
obtained by restricting the $R$-matrix partial waves to those
included in the RGM. Even compared to these results, our calculation
differs in the region of the interference between the Coulomb and the
strong interaction at small scattering angles. This discrepancy cannot
be attributed to a single partial wave, but originates from tiny
differences between the $R$-matrix phase shifts and the calculated ones.

\begin{figure*}
  \epsfig{file=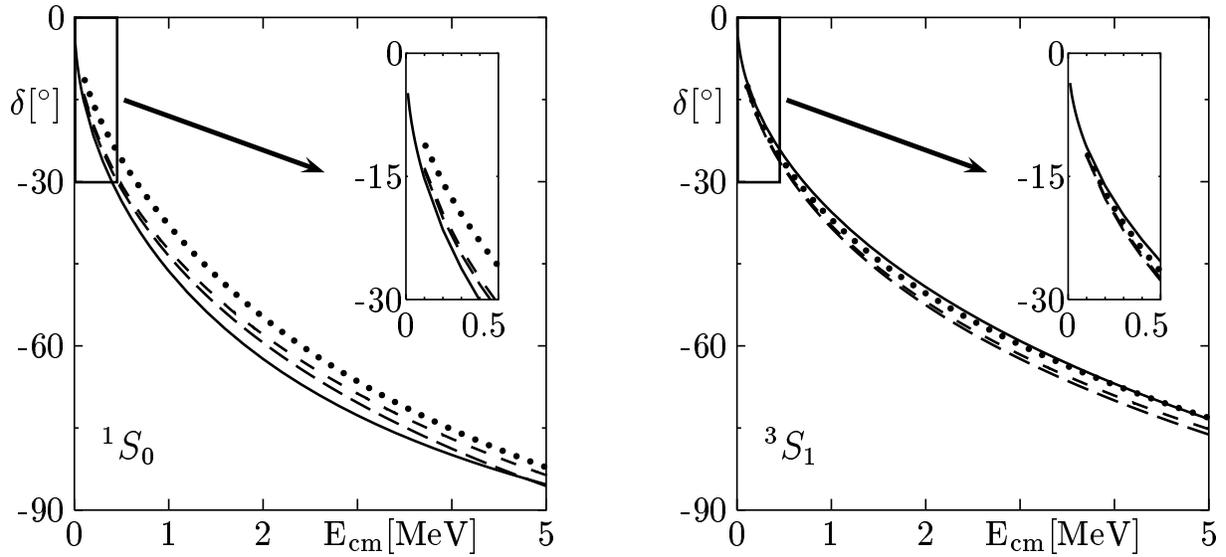, width=0.9\textwidth}
  \caption{ \label{S_NT} $S$ wave phase shifts of elastic $n - ^3$H
    scattering, the lines are as in fig.~\ref{phase_figure}. }
\end{figure*}

\begin{figure}
  \epsfig{file=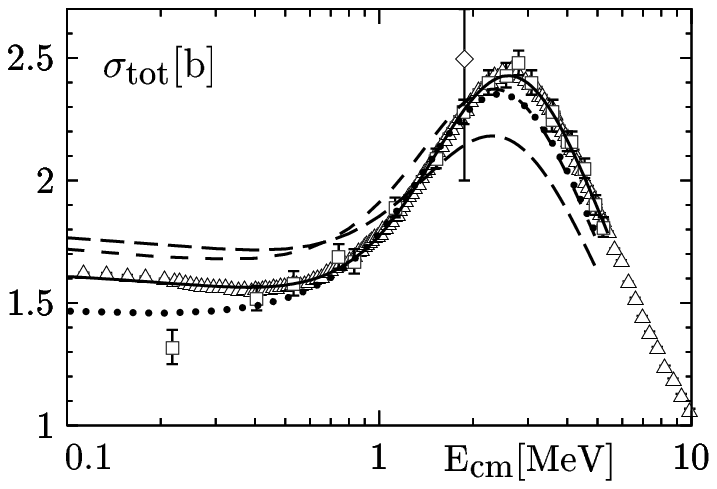, width=0.9\columnwidth}
  \caption{ \label{nt_total_xsec} $n - ^3$H total cross section.
    The experimental data is from \cite{NT_TOTAL} ($\vartriangle$),
    \cite{NT_TOTAL_BATTAT} ($\square$) and \cite{NT_TOTAL_KATSAUROV}
    ($\lozenge$), the lines are as in fig.~\ref{phase_figure}. }
\end{figure}

The total cross section of $n-^3$H scattering at low energies
depends mainly on the $^3S_1$ phase shift whose statistical weight
$2J+1$ is three compared to one for $^1S_0$. Small differences in
$^3S_1$ (fig.\ \ref{S_NT} inset) lead to the 10\% deviation
between the different models ( A$v_{18}$ dashed and long dashed,
$J_3^+$ only in fig.\ \ref{nt_total_xsec}). Note that the large
effect of the Urbana IX TNI at low energies is a combination of
the good reproduction of the low energy $^3S_1$ matrix element
and the $^1S_0$ phase shift being too small.

\begin{figure}
  \epsfig{file=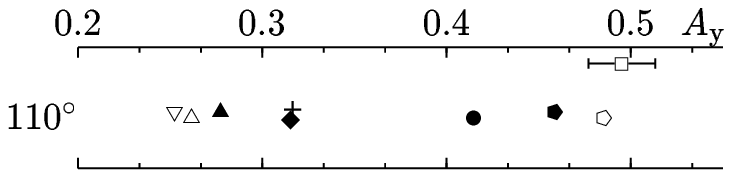, width=0.9\columnwidth}
  \caption{ \label{ay_model}
    $p - ^3$He proton analyzing power at $E_{\rm cm}=4.1$MeV and
    110 degrees for different models: A$v_{18}$ ($\bigtriangleup$),
    A$v_{18}$ $J_3^+$ only ($\bigtriangledown$), A$v_{18}$ + Urbana IX
    ($\blacktriangle$), A$v_8'$ + TLA ($\blacklozenge$), A$v_{18}$
    with $^3P_0$ ($+$) and with both $^3P_0$ and $^3P_2$ ($\bullet$)
    adjusted, $R$-matrix (pentagon) and $R$-matrix restricted to $L
    \le 2$ (filled pentagon), data point ($\square$) from
    \cite{HE3_KNUTSON}.  }
\end{figure}

As in the $\rm ^4$He system, the real challenge lies in the
polarization observables, e.\ g.\ the $p-^3$He proton analyzing
power $A_{\rm y}$. Even within the large model space our calculations
find only about half the measured $A_{\rm y}$ (fig.\ \ref{ay}) similar
to the findings in \cite{NT_FONSECA}. We know from \cite{HE4} that the
analyzing power is sensitive to the $^3P_2$ matrix element and it was
surprising that even with the $J_3^-$ distortion channels increasing
the $^3P_2$ phase shift, we cannot improve the description of $A_{\rm
  y}$. The reason is that most of the expected improvement due to the
larger $^3P_2$ matrix element is canceled by the increased $^3P_0$
phase shift that decreases $A_{\rm y}$, as can be seen when we lower
$^3P_0$ to its $R$-matrix value (fig.\ \ref{ay_model}, where the
values of $A_{\rm y}$ are displayed close to the maximum). Increasing
$^3P_2$ further to the $R$-matrix value will then raise $A_{\rm y}$
again, but still not to the experimental data.  Only a small part of
the remaining difference is due to higher partial waves that are not
included in the model space (see fig.\ \ref{ay_model}). The remainder
comes from small differences in the other partial waves.
\section{ Three-nucleon force effects }

\begin{figure}
  \epsfig{file=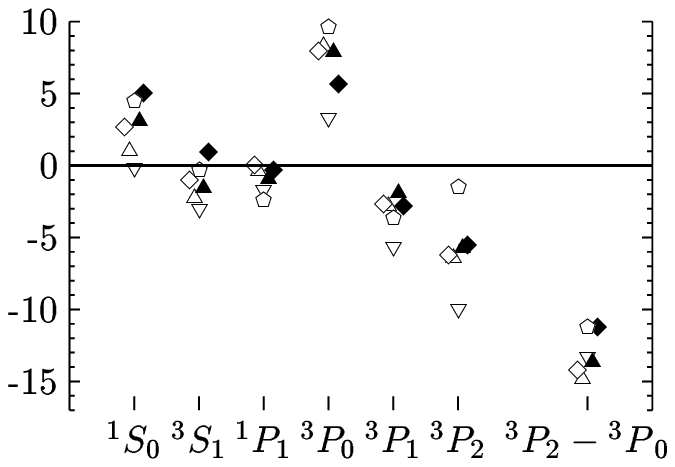, width=0.9\columnwidth}
  \caption{ \label{phase_splitting} Deviation of the calculated $p - ^3$He
    phase shifts from the $R$-matrix values in the different channels
    and the $^3P_2-{^3P}_0$ splitting for different models at $E_{\rm
      cm}=4.5 \rm MeV$: A$v_{18}$ ($\triangledown$ $J_3^+$ only,
    $\vartriangle$ full model), A$v_{18}$ + Urbana IX
    ($\blacktriangle$), A$v_8'$ ($\lozenge$) and A$v_8'$ + TLA
    ($\blacklozenge$).  }
\end{figure}

Both the A$v_8'$ and A$v_{18}$ $NN$ forces yield essentially the same
phase shifts, and we convinced ourselves that the Gaussian
parametrization of the Bonn interaction \cite{BONNFIT} also agrees
with these results.  In the previous section we showed that we cannot
reproduce the $p -^3$He analyzing power $A_{\rm y}$ mainly due to
the too small splitting of $^3P_2$ and $^3P_0$ (fig.\ 
\ref{phase_splitting}).
Therefore we included an additional TNI in our calculations. For
A$v_{18}$ we chose the $2\pi$-exchange model implemented as the
Urbana IX force \cite{AV8P_U8_U9} and for A$v_8'$ the Texas-Los Alamos
(TLA, \cite{TLA}) three-nucleon force. The latter includes only short
range operators and is claimed to resolve the $A=3$ $A_{\rm y}$
problem for a certain choice of its operator strength ($c_1=3$,
\cite{TLA}). In our calculation we chose this factor to be $c_1=1.5$
to give reasonable $\rm ^3$He binding energies in the model space
used.

In the case of the A$v_{18}$ and Urbana IX interaction, the additional
TNI leads to the expected improvement of the binding energy. For the
small $\rm ^3$He wave function we find $-6.88$ MeV instead of the
$-6.37$ MeV for A$v_{18}$ alone and the TNI increases the $D$ state
probability $P_D$ somewhat (from 7.7\% to 8.2\%). In a rather large
model space we find a binding energy of $-7.72$ MeV and kinetic energy
of $-50.05$ MeV, close to the values given in \cite{NOGGA_FAD}.

Both TNIs are unfortunately very time-consuming to evaluate and
neither of them contributes significantly to the phase shifts (fig.\ 
\ref{phase_figure}).  Especially the Urbana IX interaction does not
affect the phase shifts at all while still improving the binding
energy of $\rm ^3$He. The $^3P_2 - {^3P}_0$ splitting is somewhat
improved by the TLA force but still far from its $R$-matrix values
(fig.\ \ref{phase_splitting}).  Yet this improvement can already be
seen in the analyzing power (fig.\ \ref{ay}), where the calculation
including the TLA force gives significantly better results. Since the
Urbana force left the phase shifts almost unchanged it affects the
analyzing power (fig.\ \ref{ay}) only marginally. These effects can be
more clearly seen in fig.\ \ref{ay_model}, where the analyzing power
close to the maximum is shown for the different models.

Since the $2\pi$ exchange operators leave the scattering phase shifts
unchanged, we considered the $V_3^*$ operators proposed in
\cite{V3_SCHADOW} in addition to the Urbana IX interaction. The
$V_3^*$ uses a pion-exchange between the third particle and the center
of mass of the $NN$ system, which in turn uses the full $N-N$ interaction,
but only for $T=1$ and $S=1$. The $V_3^*$ therefore acts predominantly
in the $P$-waves. In order to study the effects of this interaction
within a relatively simple operator structure, we approximated the full
$t$-matrix by using the central part of the Argonne $v_{18}$ potential as
the $NN$-force part in $V_3^*$, and chose its strength to leave the
$\rm^3$H binding energy unchanged while reproducing the $^3P_2$ matrix
element at $E_{\rm cm}=2.4$ MeV.

The full model of the $V_3^*$ NNN interaction has only a small
influence on the binding energy of $\rm^3$H \cite{V3_AY}, which
also holds for $^4$He with the $V_3^*$ potential in our calculation.
However, these operators have a large effect on the $P$-wave phase
shifts. If we choose the strength of $V_3^*$ so that the $^3P_2$ matrix
element is reproduced and the $\rm^3$H binding remains unchanged, the
$^3P_0$ phase shift unfortunately increases again, and therefore part of
the improvement due to the larger $^3P_2$ matrix element is cancelled.
Nevertheless, Urbana IX and $V_3^*$ together achieve an $A_{\rm y}$ of
0.35 at $E_{\rm cm}=4.1$ MeV, a much larger effect than the Urbana IX
force alone achieved.

\section{Conclusions}

We have discussed a new $R$-matrix analysis of the currently available
experimental data. The phase shifts calculated in this analysis were
compared to an RGM calculation of $p - ^3$He scattering. We showed
that realistic $NN$ interactions describe most of the phase shifts
quite well but fail to reproduce the $^3P_2$ and $^3P_0$ phase shifts.
The calculated splitting between these two channels is much too small, and
neither the Urbana IX nor the TLA three-nucleon force is able to improve
the splitting significantly.  In fact, it is more important to include in
the calculation  negative parity states of the three-nucleon subsystem
than one of these two TNIs. These findings show that new contributions to
the $NNN$ force acting on the $P$-waves should be considered, like an
$LS$ type TNI, as proposed in \cite{kievsky_LS} for the $N-d$ analyzing
powers, or the $V_3^*$ operators proposed in \cite{V3_SCHADOW}.

\begin{acknowledgements}
  This work is supported by the DFG ({\em Graduiertenkolleg
  Erlangen-Regensburg}) and the BMBF (contract 06ER926) and used
  resources at several computer centers (RRZE Erlangen, NIC J\"ulich,
  SSC Karlsruhe and LRZ M\"unchen).  We want to thank G.\ Wellein and
  G.\ Hager at the RRZE for their help.  The U.~S. Department of Energy
  supported the work of G.~M.~H. on this study.
\end{acknowledgements}

\bibliography{article.bib}

\end{document}